\documentclass[11pt]{article}
\usepackage{hyperref}
\pdfoutput=1

\begin{document}

\title{Shape oscillation of bubbles in the acoustic field}

\author{Keishi Matsumoto$^{1}$, Ichiro Ueno$^{2}$ \\
\\\vspace{0pt} $^{1}$Graduate School, Tokyo University of Science, \\\vspace{6pt} 2641 Yamazaki, Noda, Chiba 278-8510, JAPAN
\\\vspace{6pt} $^{2}$Tokyo University of Science, Noda, JAPAN}

\maketitle


\begin{abstract}
The authors introduce dynamics of multiple air bubbles exposed to
ultrasonic wave while ascending in water in the present fluid dynamics
video. The authors pay attention to the shape oscillation
and the transition from the volume to the shape oscillations of the
bubble. Correlation between the bubble size and the mechanism
of the excitation of the shape oscillation is introduced.
\end{abstract}

\section{Introduction}


The video is
\href{http://ecommons.library.cornell.edu/bitstream/1813/14088/2/APS_Keishi_Matsumoto.mpg}{Video1}.

To understand a behavior of the bubbles, their shape
and volume oscillations, under periodic external force is of one of
great importance in order to control heat/mass transfer in multiphase
flows. A large number of researches have been carried out on the volume
oscillation by theoretical and experimental approaches (reviewed by
Plesset \& Prosperetti (1) and Feng \& Leal (2)). On the shape
oscillation, however, there exist few experimental works although there
have been accumulated theoretical predictions. In the present study, The
authors introduce the effect of the preceding bubble in the acoustic
field on the shape oscillation arisen on the following bubble.

REFERENCES
\\(1) Pleset, M. S. Prosperetti,A., Bubble Dynamics and Cavitaion,Ann. Rev. Fluid Mech.,29(1977),145-185
\\(2) Feng, Z. C. \& Leal, L. G., Nonlinear Bubble Dynamics, Ann. Rev. Fluid Mech., 29(1997), 201-243
%
\end{document}